\documentclass{bioinfo}

\usepackage{url}

\usepackage{bm} 

\def\Fun#1{\ensuremath{\textup{#1}}} 

\usepackage{xspace}
\newcommand{\Tool}{SBML2\-Mod\-el\-ica\xspace}
\newcommand{\ToolUrl}{\url{https://bitbucket.org/mclab/sbml2modelica}}
\newcommand{\SBMLLV}[2]{SBML Level #1 Version #2\xspace}
\newcommand{\SBMLLatestStandard}{\SBMLLV{3}{2}}
\newcommand{\OpenModelicaVersion}{v1.32.2\xspace}
\newcommand{\JModelicaVersion}{v2.4\xspace}
\newcommand{\pre}[1]{\mathrm{pre}\left(#1\right)}
\newcommand{\vecvar}[1]{\ensuremath{\bm{#1}}}

\newcommand{\derive}[1]{\ensuremath{\frac{\mathrm{d} #1}{\mathrm{d} t}}}
\newcommand{\eg}{\textit{e.g., }}
\newcommand{\ie}{\textit{i.e., }}
\newtheorem{example}{Example}
\newcommand{\Stoich}[1]{\nu(#1)}

\def\FigureStandardWidth{0.9\linewidth}

\newcommand{\ModelicaCompartmentSize}[1]{\ensuremath{#1^{\Fun{size}}}}
\newcommand{\ModelicaSpeciesAmount}[1]{\ensuremath{#1^{\Fun{am}}}}
\newcommand{\ModelicaSpeciesConc}[1]{\ensuremath{#1^{\Fun{con}}}}

\newcommand{\TestSuiteNbAll}{1623\xspace}
\newcommand{\TestSuiteNbIncluded}{1588\xspace}
\newcommand{\TestSuiteNbPasses}{1532\xspace}
\newcommand{\TestSuiteNbFails}{56\xspace}
\newcommand{\TestSuitePassPerc}{96.47\%\xspace}

\newcommand{\TestSuiteNbFailsNotSignificant}{8\xspace}

\newcommand{\TestSuiteNbFailsUnsupportedCombs}{48\xspace}

\newcommand{\BioModelsNbAll}{641\xspace}
\newcommand{\BioModelsNbIncluded}{613\xspace}
\newcommand{\BioModelsNbVarsMin}{6\xspace}
\newcommand{\BioModelsNbVarsMax}{5438\xspace}
\newcommand{\BioModelsTimeout}{360 seconds\xspace}

\newcommand{\BioModelsManualNbVarsMin}{250\xspace}
\newcommand{\BioModelsHorizon}{100\xspace}
\newcommand{\BioModelsMaxTimeStep}{0.01\xspace}

\let\cite\citep

\begin{document}
\copyrightyear{2020} 
\pubyear{2020}

\access{Volume 36, Issue 7, 1 April 2020, Pages 2165–2172}
\appnotes{Original Paper}

\firstpage{2165}

\subtitle{Systems biology}

\title[\Tool]{\Tool: 
    Integrating biochemical models within open-standard simulation ecosystems
}
\author[Maggioli \textit{et~al}.]{
    F.~Maggioli\,$^{\text{\sfb 1}}$,
    T.~Mancini\,$^{\text{\sfb 1}\ast}$,
    E.~Tronci\,$^{\text{\sfb 1}}$
}

\address{$^{\text{\sf 1}}$Computer Science Department, Sapienza University of Rome, Italy%
}

\corresp{$^\ast$To whom correspondence should be addressed.}

\history{}

\editor{}

\abstract{%
	\textbf{Motivation:}
	SBML is the most widespread language for the definition of biochemical models.
	Although dozens of SBML simulators are available, there is a general lack of support to the integration of SBML models within open-standard general-purpose simulation ecosystems. This hinders co-simulation and integration of SBML models within larger model networks, in order to, \eg enable \emph{in-silico clinical trials} of drugs, pharmacological protocols, or engineering artefacts such as biomedical devices against Virtual Physiological Human models.
	\\
	\emph{Modelica} is one of the most popular existing open-standard general-purpose simulation languages, supported by many simulators. Modelica models are especially suited for the definition of complex networks of heterogeneous models from virtually all application domains. 
	Models written in Modelica (and in 100+ other languages) can be readily exported into \emph{black-box Functional Mock-Up Units (FMUs)}, and seamlessly co-simulated and integrated into larger model networks within \emph{open-standard language-independent simulation ecosystems}.
	\\
	\textbf{Results:}
	In order to enable SBML model integration within heterogeneous model networks, we present \Tool, a software system translating SBML models into well-structured, user-intelligible, easily modifiable Modelica models. \Tool is \SBMLLatestStandard--compliant and succeeds on \TestSuitePassPerc of the SBML Test Suite Core (with a few rare, intricate, and easily avoidable combinations of constructs unsupported and cleanly signalled to the user).
	Our experimental campaign on \BioModelsNbIncluded models from the BioModels database (with up to \BioModelsNbVarsMax variables) shows that the major open-source (\emph{general-purpose}) Modelica and FMU simulators achieve performance comparable to state-of-the-art \emph{specialised} SBML simulators.
	\\
	\textbf{Availability and Implementation:}
	\Tool is written in Java and is freely available for non-commercial use at \mbox{\ToolUrl}
	\\
	\textbf{Contact:} T.\ Mancini, \href{mailto:tmancini@di.uniroma1.it}{tmancini@di.uniroma1.it}
}

\maketitle

\section{Introduction}
\label{sec:intro}
The mathematical modelling and simulation of biochemical systems is of paramount importance in several areas, \eg computational and systems biology, model-based pharmacology, chemistry.
The current \emph{de facto} standard for \emph{modelling} biochemical systems is \emph{Systems Biology Markup Language} (SBML, \citealp{hucka-etal:2003:bioinf}, \mbox{\url{http://www.sbml.org}}), an XML-based markup language allowing the definition of biochemical models in terms of reactions, species, compartments, and parameters. 
SBML allows the quantitative modelling of various kinds of biological phenomena, including metabolic networks, cell signalling pathways, regulatory networks, infectious diseases, just to mention a few.

\emph{Simulation} of SBML models of practical relevance is crucial for their analysis, as they are often too large or intricate for being analysed statically. 
Indeed, many third-party simulators of SBML models have been developed and are currently publicly available (see, \eg the SBML web-site).

\subsection{Motivations}
\label{sec:motivations}
Available SBML simulators do not fully support the integration, within open-standard simulation ecosystems, of SBML models with models defined using \emph{other} languages.
This severely hinders the possibility to \emph{co-simulate} and \emph{integrate} SBML models within large \emph{model networks} comprising biochemical as well as other kinds of models, possibly at different levels of abstraction (\emph{multi-scale} model networks, see, \eg \citealp{debono-etal:2012:biotech}), and applying standard systems engineering approaches for the model-based analysis of such heterogeneous model networks.

For example, the interconnection of quantitative models of the human physiology (\eg Physiomodel, \citealp{matejak-etal:2015:embc}), drugs pharmacokinetics/pharmacodynamics (\eg Open Systems Pharmacology Suite, \citealp{eissing-etal:2011:frontiers}), (possibly semi-autonomous) biomedical devices, pharmacological protocol guidelines or treatment schemes, enables the set-up of \emph{in silico clinical trials} for the (model-based) safety and efficacy pre-clinical assessment of such drugs, protocols, treatments, devices, using \emph{standard system engineering approaches} to perform their simulation-based analysis at system level (see, \eg \citealp{kanade-etal:2009:cav,mancini-etal:2013:cav,mancini-etal:2014:pdp,zuliani-etal:2013:fmsd,zuliani:2015:jttt,mancini-etal:2016:micpro,mancini-etal:2017:ipl}).
Works in this direction include, \eg \cite{schaller-etal:2016:mpc,messori-etal:2018:individualized}, where a model-based verification activity of a sensor-augmented insulin pump is conducted against a model of the human glucose metabolism in patients with diabetes mellitus, 
\cite{madec-etal:2019:lab-on-chips}, where a model of a penicillin bio-sensor (integrating biochemistry, electrochemistry, and electronics models) is simulated to compute a first dimensioning of the sensor, 
and \cite{mancini-etal:2014:fmcad,mancini-etal:2015:iwbbio}, where representative populations of virtual patients are generated from parametric models of the human physiology, a key step to enable \emph{in silico} clinical trials (see, e.g., \citealp{mancini-etal:2018:rcra-treatment}).

One of the most widely adopted open-standard languages for modelling dynamical systems is Modelica (\url{http://www.modelica.org}), a general-purpose fully-fledged language based on ordinary differential as well as algebraic equations plus procedural snippets. 
The language supports object-orientation and allows the definition of complex systems as networks of smaller subsystems.

Modelica is widespread in application domains as diverse as mechanical, electrical, electronic, hydraulic, thermal, control, electric engineering, but also physiology and pharmacology (see, \eg \citealp{matejak-etal:2015:embc}), and several efficient and highly-configurable simulators are currently available: proprietary (\eg Dymola 
and Wolfram System Modeler) 
as well as open-source (\eg OpenModelica and JModelica). 

A Modelica model can also be easily exported into a Functional Mock-Up Unit (FMU), an executable \emph{opaque} (binary) object implementing the Functional Mock-Up Interface (FMI, \url{http://fmi-standard.org}), one of the currently most widespread open standards for model exchange, integration and co-simulation. 
Being \emph{black-box}, FMU models can be shared or integrated within larger model networks while protecting their \emph{intellectual property} (see, \eg \citealp{mancini-etal:2016:fundam}). This is crucial when sharing, integrating, or co-simulating models coming from different providers (\eg pharma companies, or manufacturers of novel biomedical devices).
The FMI/FMU standard is currently supported by more than 100 simulators for \emph{virtually all} application domains, making it the largest open-standard ecosystem for (language-independent) model exchange, integration, and co-simulation.

\subsection{Contributions}
\label{sec:contributions}
In this paper we present \Tool, a software system that translates SBML models into well-modularised user-intelligible \emph{Modelica} code, which preserves both the structure and the documentation of the input SBML models. 
The generated Modelica models can then be easily modified, integrated within other models, and can be readily run using \emph{any} available Modelica simulator. Furthermore, the generated Modelica models can be easily exported into FMUs, thus allowing their seamless co-simulation and integration into model networks within open-standard language-independent simulation ecosystems (a helper tool is provided in the \Tool repository which generates a FMU directly from an SBML model by leveraging the JModelica API).

\Tool is compliant to the \emph{latest} SBML standard (\SBMLLatestStandard, \citealp{hucka-etal:2018:sbml-l3v2}) and succeeds on \TestSuitePassPerc of the SBML Test Suite Core v3.3.0 (see Section~\ref{sec:results:correctness}), with a few rare, intricate, and easily avoidable combinations of constructs (see Section~\ref{sec:discussion}) unsupported and cleanly signalled to the user.
Furthermore, our experimental campaign on \BioModelsNbIncluded models from the BioModels database (with up to \BioModelsNbVarsMax variables) shows that major open-source (\emph{general-purpose}) Modelica and FMU simulators (OpenModelica and JModelica, with the latter that converts the input Modelica model into an FMU and then simulates such an FMU), when used in their default configurations achieve performance comparable to state-of-the-art \emph{specialised} SBML simulators (see Section~\ref{sec:results:performance}). 

\Tool can be freely downloaded for non-commercial uses. 
The system has been implemented in Java and can be executed on all platforms for which a Java Virtual Machine is available. This includes most computer operating systems.

\subsection{Available SBML simulators}
\label{sec:soa}
A plethora of systems for the simulation of models written in SBML are currently available (most of them are listed in the SBML web-site), 
and a comprehensive review of them is out of the scope of this paper.
We note, however, that both their functionalities and compliance to the SBML standard is highly variable. 

In particular, only for six systems a certified report of their compliance to the official SBML Test Suite Core is (at the time of writing) publicly available (see the SBML web-site). 
Some of such systems (namely: libRoadRunner, \citealp{somogyi-etal:2015:libroadrunner}; libSBMLSim, \citealp{takizawa-etal:2013:libsbmlsim}; and Simulation Core Library, \citealp{keller-etal:2013:simcorelib}) are pure SBML simulators, allowing the user to numerically simulate the SBML model given as input. All of them support a previous SBML standard (\SBMLLV{3}{1}), while \Tool supports the \emph{latest} standard (\SBMLLatestStandard) with only a few minor limitations (see Section~\ref{sec:discussion}).

The other systems (namely: BioUML, \citealp{kolpakov:2019:biouml}; iBioSim, \citealp{myers-etal:2009:ibiosim}; and COmplex PAthways SImulator --COPASI, \citealp{lee-etal:2006:copasi}) are more general platforms that, beyond model simulation, allow the user to modify, extend, and connect different SBML models together. Of them, only BioUML supports \SBMLLatestStandard, as iBioSim and COPASI only support the older \SBMLLV{3}{1}.

By translating SBML models into an open-standard general-purpose widely-adopted simulation language as Modelica (preserving both the structure and the documentation of the input models), \Tool not only allows simulation of the generated models (using any Modelica simulator), but also opens up a huge plethora of new possibilities to integrate SBML biochemical models with models of other kinds of systems (see Section~\ref{sec:motivations}) written in languages different than SBML.

Enabling interoperability and integration of biochemical models into cross-domain model networks has been strongly advocated. 
Attempts in this directions include, \eg \cite{madec-etal:2017:bbspice}, where biochemical models are converted into Spice, a standard integrated electronic circuit simulator, for the model-based design of bio-sensors and labs-on-chip. Model conversion is performed exploiting clever analogies between the behaviour of biochemical systems and electronic circuits and between molecular diffusion and heat diffusion \cite{gendrault-etal:2014:ieee-tbe,madec-etal:2019:lab-on-chips}. 
\Tool acts at a higher level, by translating SBML models into a genuinely general-purpose cross-domain open system modelling language (Modelica), hence enabling seamless integration and co-simulation of SBML models with models of virtually \emph{all} application domains, without the need to exploit cross-domain analogies, hence fully preserving model readability and extensibility. 
The possibility to export Modelica models into FMUs is one step further, allowing model integration and co-simulation in a language-independent way.

Previous approaches to translate SBML models into \emph{general-purpose} simulation platforms include
the SimBiology Matlab toolbox and 
Wolfram SystemModeler (a proprietary simulator accepting the Modelica language) with the BioChem plug-in \cite{larsdotter-etal:2003:biochem,fritzson-etal:2007:biochem}. 
In particular, by providing a user-friendly interface and high-level library abstractions, the BioChem SystemModeler plug-in allows the definition of visually appealing biochemical networks in some Modelica editors. 
Differently from \Tool, both SimBiology and Syst\-emMod\-eler+BioChem are based on commercial simulators and support only subsets of older SBML standards (\SBMLLV{3}{1} and \SBMLLV{2}{4} respectively), with several major limitations, including lack of support of delayed and prioritised events.
Also, for none of them a compliance assessment to the SBML Test Suite core is available.
%

\begin{methods}
\section{Materials and methods}
\label{sec:methods}
In the following, we briefly outline the main structure of an SBML (Section~\ref{sec:methods:sbml}) and of a Modelica model (Section~\ref{sec:methods:modelica}), before sketching (Section~\ref{sec:methods:encoding}) how \Tool generates a structured Modelica model from an input SBML model.

\subsection{High-level view of an SBML model}
\label{sec:methods:sbml}
Here we recall the main constructs of SBML, namely: parameters, compartments, species, and events.
The reader interested to a more in-depth description is referred to the official SBML web-site (\url{http://www.sbml.org}) for the full language specification.

\emph{Parameters} 
denote quantities with a symbolic name. Such quantities can be either constant or varying during model evolution.

\emph{Compartments}
denote containers of a particular type and positive \emph{size} (possibly varying during model evolution).

\emph{Species} 
represent model entities (\eg biochemical substances), whose \emph{amount} may vary during model evolution.
Each species belongs to a compartment.
Species may take part to \emph{reactions}. At any time, the \emph{concentration} of a species in its compartment is defined as $\frac{\Fun{amount}}{\Fun{size}}$, where \Fun{size} is the size of the species compartment at that time.

Model parameters, species, and size of compartments are defined by means of \emph{model variables} and can be assigned to values.
An \emph{initial assignment} defines the value of a model variable at time 0.

\emph{Reactions} 
are statements describing any transformation, transport or binding process that changes the amount of one or more species. 
A reaction of the form $\alpha \to \beta$ (where $\alpha$ and $\beta$ are \emph{mixtures}, defined as linear combinations of species, \eg $\alpha = s_1 + 2 s_2$, $\beta = s_3$, where $s_1, s_2, s_3$ are species) describes how (and how much of) certain species (those in $\alpha$, called reactants) are transformed into certain other species (those in $\beta$, called products). Reactions have associated \emph{kinetic rate expressions} that describe how quickly they take place. 

According to the SBML specification, for any species $s$, the set of reactions $R_1, \ldots, R_n$ in which $s$ occurs (together with their associated kinetic rate expressions $k_{R_1}, \ldots, k_{R_n}$) collectively define the time derivative of the available amount of $s$ as: 
\begin{equation}
\label{eq:methods:der_species}
\derive{s} = 
\sum_{i=1}^{n} k_{R_i} \times \Stoich{s, R_i}
\end{equation}
where $\Stoich{s, R_i}$ is the sum of the coefficients that multiply the occurrences of $s$ in reaction $R_i$. Coefficients occurring on the left side of $R_i$ are multiplied by $-1$ in order to model species consumption, while those occurring on the right side are taken as they are in order to model species production (see forthcoming Example~\ref{ex:methods:example}).

\emph{Events} 
represent \emph{instantaneous} and \emph{discontinuous} changes in the value of some quantities (\eg amounts of species, parameters, size of compartments) of the model. 
An event is defined in terms of a \emph{trigger condition} (a Boolean formula), and a set of \emph{assignments}, which update some model variables when the event \emph{occurs} (\ie when the trigger condition switches from false to true).
Optionally, events can be \emph{delayed} by a certain time interval, whose length could change during model evolution.
To avoid that two events occurring at the same time assign different values to the same variable, a \emph{priority} expression (on the model variables) can be defined for events. Event priority expressions, evaluated when events occur, define the order in which concurrent events must be handled.

SBML events can be either persistent or non-persistent.
Let $e$ be an event, $t$ be the time instant when the trigger condition of $e$ becomes true, and $d$ be the event delay.
\emph{Non-persistent} (respectively, \emph{persistent}) event $e$ must be executed at time instant $t+d$ \emph{only if} (respectively, \emph{regardless of whether}) the trigger condition remains true during the whole delay period (\ie from time $t$ to time $t+d$).

\emph{Rules} provide additional means to define the values of variables in a model in ways that cannot be expressed using reactions or initial assignments. The following three types of rules are provided (below, $\vecvar{V}$ is a set of --possibly all-- model variables).
\emph{Algebraic rules} are of the form $f(\vecvar{V}) = 0$.
\emph{Assignment rules} are of the form: $x = f(\vecvar{V} - \{x\})$.
Finally, \emph{rate rules} define the rate of change of a model variable and are of the form: $\derive{x} = f(\vecvar{V})$.

Example~\ref{ex:methods:example} shows a simple SBML model that will be used as a running example when outlining how \Tool works (Section~\ref{sec:methods:encoding}). Although the model is clearly artificial and might not recall a known biochemical mechanism, it has the merit of compactly showing all the most important SBML constructs that we will address in the remainder of the paper.

\begin{example}[Running example]
    \label{ex:methods:example}
    \newcommand{\Header}[1]{\par\smallskip\noindent\textup{\textbf{#1}}}
    Our SBML model (whose code is available in the \Tool repository) consists of the following elements:
    
    \Header{Parameters:}
    $p_1$ with constant value 
    $10^{-6}~[\text{l}~\text{sec}^{-1}]$;
    $p_2$ whose value is initially set to 
    $1~[\text{mol}^{-1} \text{sec}^{-1}]$;
    $p_3$ with constant value 
    $10^{-3}~[\text{mol}]$;
    $p_4$ whose initial value is set to 
    $300~[\text{sec}]$.
    
    \Header{Species}~(all in $[\text{mol}]$):	
    $s_1$, $s_2$, $s_3$ (initially set to $10^{-3}$),
    and
    $s_4$. 
    
    \Header{Compartments:} One compartment $c$ containing all the species.
    
    \Header{Reactions:} Reaction
    $R : s_1 + 2 s_2 \longrightarrow s_3$ 
    with kinetic rate expression 
    $k_R = p_2 s_1 s_2~[\text{mol}~\text{sec}^{-1}]$.
    
    \Header{Events:}
    \begin{itemize}
        \item $e_1$ with trigger condition $s_1 s_2 \leq 10^{-7} \lor s_3 s_4 \leq 10^{-7}$ and priority equal to $s_4$. When the event is triggered, parameter $p_2$ is set to $0$.
        \item $e_2$ with the same trigger condition as $e_1$, but with a delay of $p_4$ and priority equal to $s_2$. When triggered, parameter $p_2$ is set to $-1$ and parameter $p_4$ is set to $0$.
    \end{itemize}

    \Header{Rules:}
    \begin{itemize}
        \item Rate rule $r_1$, defining $\derive{p_2} = 0$;
        \item Assignment rule $r_2$, which sets the $\Fun{size}$ of compartment $c$ to $1 + p_1 \times t~[\text{l}]$, where $t$ is the value of the current time-instant;
        \item Algebraic rule $r_3$, which imposes that constraint
        $\frac{s_2}{s_1} - \frac{s_3}{s_4} = 0$
        holds at all time points.
    \end{itemize}
    
    \Header{Initial assignment} which sets the value of $s_2$ to $p_3$ at time zero.
\end{example}

The model in Example~\ref{ex:methods:example} comprises 4 species ($s_1, \ldots, s_4$) all belonging to a single compartment (whose size is constantly increasing as dictated by assignment rule $r_2$). The time-evolution of the available amount of $s_1$, $s_2$, and $s_3$ is governed by reaction $R$, while that of $s_4$ is determined by algebraic rule $r_3$ (hence, $s_4$ is always equal to $\frac{s_1 s_3}{s_2}$).
The kinetic rate expression $k_R$ of reaction $R$ dictates how quickly the reaction takes place. 
In our example, reaction $R$ defines the following time-derivatives for the available amounts of the involved species, according to Eq.~\eqref{eq:methods:der_species}:
$\derive{s_1} = -k_R$, $\derive{s_2} = -2 k_R$, $\derive{s_3} = k_R$.
However, $k_R$ is not constant in time, and, moreover, depends on parameter $p_2$, whose value is affected by the two events $e_1$ and $e_2$. This makes our model not trivial.

Figure~\ref{fig:methods:example} shows the time evolution of the available amount of each species of our model, when starting from the initial state, where the available amount of each species is $10^{-3}$, $p_1 = 10^{-6}$, $p_2 = 1$, $p_3 = 10^{-3}$, $p_4 = 300$.
With $p_2 = 1$, $k_R$ is positive, hence $R$ defines a reaction where $s_1$ and $s_2$ are consumed in favour of the production of $s_3$.
When, at time $1233.10~[\text{sec}]$, $s_1 s_2$ becomes $\leq 10^{-7}$ both events $e_1$ and $e_2$ are triggered. Given that $e_2$ has a \emph{delay} of $p_4 = 300~[\text{sec}]$, $e_1$ is processed immediately, while $e_2$ is processed only after $300$ more seconds.
This implies that $p_2$ is immediately set to 0 (as dictated by $e_1$). The kinetic rate expression $k_R$ of reaction $R$ is thus set to $0$ and the system stabilises.
When, at time $1533.10~[\text{sec}]$, also $e_2$ is processed, parameter $p_2$ is set to $-1$ and $p_4$ to $0$.
The new value for $p_2$ makes $k_R$ negative, hence reaction $R$ turns into modelling the consumption of $s_3$ in favour of the production of $s_1$ and $s_2$. 
This behaviour continues until time $3078.48~[\text{sec}]$, when $s_3 s_4$ becomes $\leq 10^{-7}$, thus triggering again both $e_1$ and $e_2$. This time, however, $e_2$ has a delay of $p_4 = 0~[\text{sec}]$, hence $e_1$ and $e_2$ are now triggered and processed \emph{simultaneously}. Since the priority of $e_2$ (\ie $s_2$) is higher than the priority of $e_1$ (\ie $s_4$), the SBML specification stipulates that the two events are processed in the order $e_2, e_1$. This implies that $p_2$ is first set to $-1$ and then (at the \emph{same} time point) to $0$. 
With $p_2$ being set to $0$, also $k_R$ becomes $0$ and the system stabilises again.

\begin{figure}
    \begin{center}
        \includegraphics[width=\FigureStandardWidth]{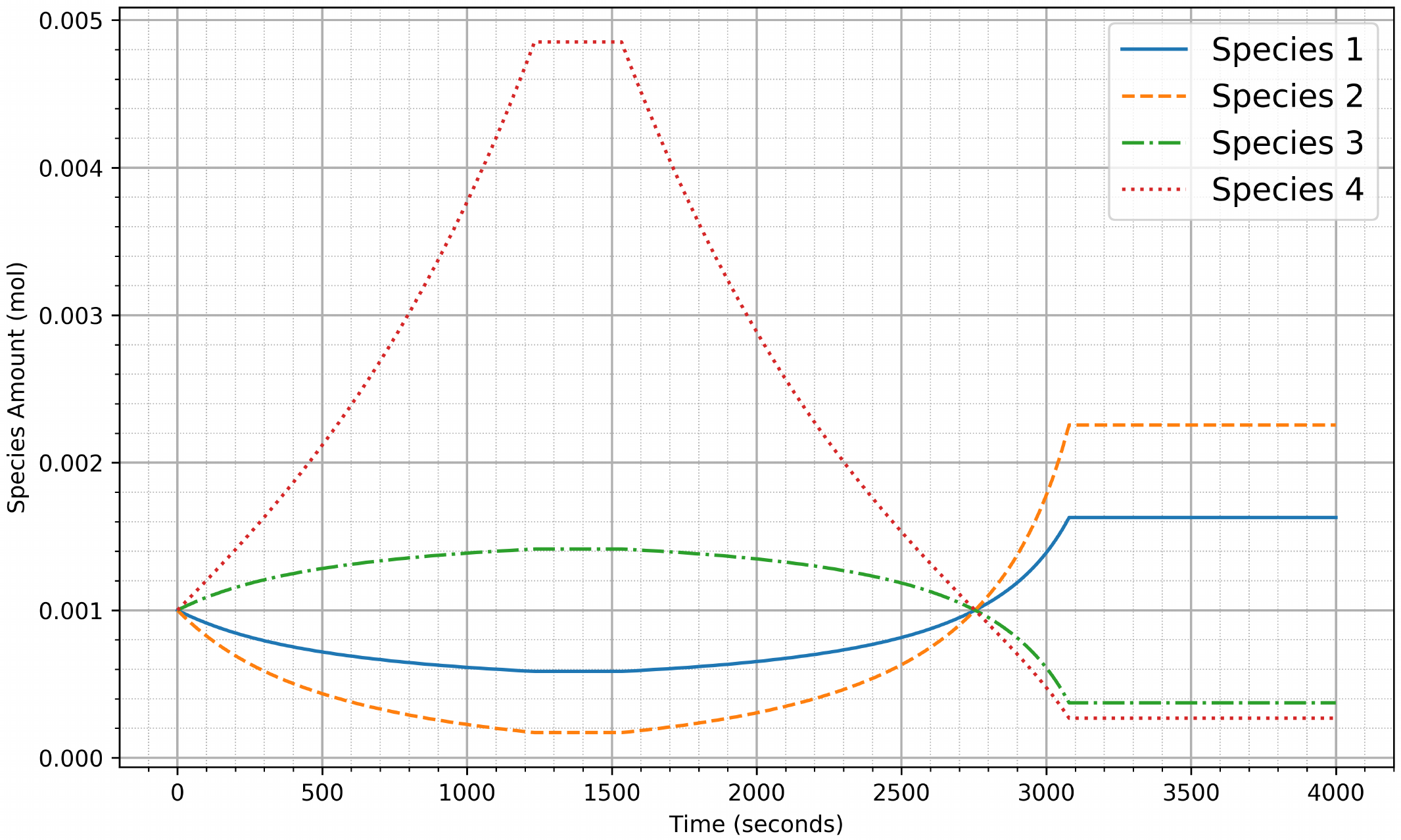}	
        
        \caption{Time evolution of the SBML model of Example~\ref{ex:methods:example}.
        }
        \label{fig:methods:example}    
    \end{center}    
\end{figure}

\subsection{High-level structure of a Modelica model}
\label{sec:methods:modelica}
Modelica is an object-oriented language for the definition of systems of differential-algebraic equations. Below, we briefly recall the general structure of a Modelica model. 
The reader interested to a more in-depth description is referred to the official Modelica web-site (\url{http://www.modelica.org}) for the full language specification.

A Modelica model is a network of \emph{objects}. 
Each object defines a set of \emph{variables}, initial assignments as well as differential and algebraic \emph{equations} for them, \emph{events} and \emph{algorithmic} sections.
At any time point, the state of a model is the value of the variables belonging to all its objects.
Variables belonging to an object can be referenced from other objects via proper \emph{connections}.

\subsection{Modelica code generation}
\label{sec:methods:encoding}
Differently from the BioChem plug-in of Wolfram SystemModeler (the only other Modelica-based SBML simulator available), \Tool does not rely on library abstractions, but generates stand-alone yet well-structured and human-intelligible Modelica code. 

In particular, the Modelica model generated by \Tool starting from the SBML model given as input is a network of different objects of 5 different classes (whose code is stored in separate files), following the structure shown in Figure~\ref{fig:methods:encoding:structure}.

This structure ensures full portability and extensibility of the generated Modelica code (no plug-ins are required), and enables easy modifications at the level of each basic component (as no library classes are involved). 

\begin{figure}
    \begin{center}
        \includegraphics[width=\FigureStandardWidth]{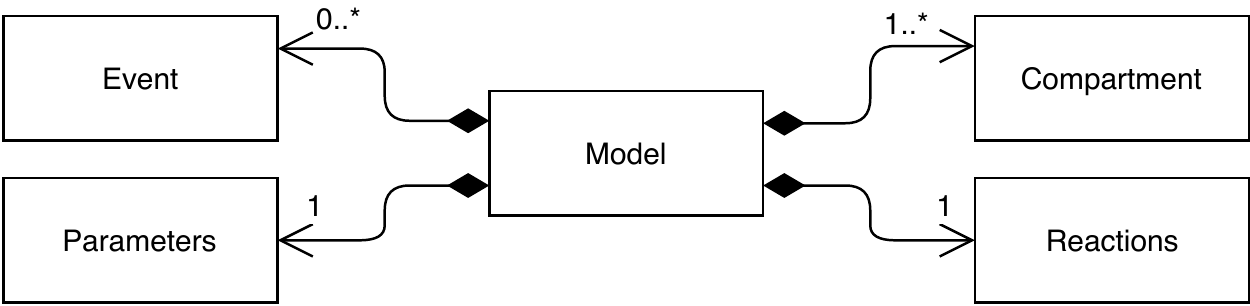}
        
        \caption{
            UML class diagram of the Modelica models generated by \Tool.
        }
        \label{fig:methods:encoding:structure}
    \end{center}        
\end{figure}

\subsubsection{The Model object}
\label{sec:methods:encoding:model}
The Model object acts as an orchestrator, by holding links to the other model objects and defining, via proper connections, the inter-object visibility of model variables.

Furthermore, the Model object defines the algebraic equations encoding the algebraic rules occurring in the input SBML model, as they may constrain variables belonging to different Modelica objects.
For instance, the Model object generated from Example~\ref{ex:methods:example} would define an algebraic equation encoding algebraic rule $r_3$.

Finally, the Model object hosts a set of auxiliary functions (also generated by \Tool) needed to handle conflicting assignments from simultaneous events (see Section~\ref{sec:methods:encoding:event}).

\subsubsection{The Compartment objects}
\label{sec:methods:encoding:compartment}
\Tool defines a Modelica \emph{Comp\-art\-ment} object for each SBML model compartment. Such objects are then linked with the Model object.

The object variables define the compartment size and the amount and concentrations of all the species belonging to the compartment.
For example, the object associated to compartment $c$ of Example~\ref{ex:methods:example} would define variables 
\ModelicaCompartmentSize{c}, 
\ModelicaSpeciesAmount{s_1}, 
\ModelicaSpeciesConc{s_1},
\ldots,
\ModelicaSpeciesAmount{s_4}, 
\ModelicaSpeciesConc{s_4}.
Initial assignments to the object variables (if defined within the input SBML model, as it happens in Example~\ref{ex:methods:example}) and differential/algebraic equations for the species belonging to the compartment as well for the compartment size are encoded using information from SBML initial assignments, reactions, and rules. 

For instance, the Modelica code generated from Example~\ref{ex:methods:example} would define 
the time-derivative of the amount of each species involved in reaction $R$ (\ie of variables \ModelicaSpeciesAmount{s_1}, \ModelicaSpeciesAmount{s_2}, and \ModelicaSpeciesAmount{s_3}) 
as stipulated by Eq.~\eqref{eq:methods:der_species}, referencing variables (which store data on the reaction) belonging to the Reactions object (see below).
Hence, we would have differential equations 
$\derive{\ModelicaSpeciesAmount{s_i}} =
\Stoich{s_i, R} \times k_R
$, for $i \in [1,3]$. 
Also, algebraic equation 
$\ModelicaCompartmentSize{c} = 1 + p_1 \times t$
(where $t$ refers to the current time instant) would be generated to encode assignment rule $r_2$.
The time-derivative of the amount of any species involved in a rate rule (there are none in Example~\ref{ex:methods:example}) would instead be encoded using its associated rule.

Conversely, equations for species whose amount is defined by means of an SBML algebraic rule (like $s_4$ in Example~\ref{ex:methods:example}) are defined within the Model object, as they represent constraints whose scope may span several Modelica objects.

Finally, variables representing the concentrations of all species are defined from their amounts. 
So, for compartment $c$ of Example~\ref{ex:methods:example}, we would have
variable assignments
$
\ModelicaSpeciesConc{s_i} \gets 
\frac
{\ModelicaSpeciesAmount{s_i}}
{\ModelicaCompartmentSize{c}}
$ 
($i \in [1,4]$).

Suitable assertions are injected into the Modelica code to ensure that variables referring to compartment sizes are always strictly positive (as dictated by the SBML semantics), hence guaranteeing that species concentration variables are always defined.

\subsubsection{The Reactions object}
\label{sec:methods:encoding:reactions}
\Tool defines a single \emph{Reactions} Modelica object storing data for all reactions defined in the input SBML model.
Such an object is then linked with the Model object.

Object variables hold the kinetic rate expression for each reaction $R$ in the model, as well as the coefficients $\Stoich{s,R}$ for each species $s$ occurring in reaction $R$ (see Eq.~\eqref{eq:methods:der_species}).
Thus, as for the single reaction $R$ of Example~\ref{ex:methods:example}, we would have variable
$k_R$ defining the kinetic rate expression of $R$ via the algebraic equation $k_R = p_2 \times \ModelicaSpeciesAmount{s_1} \times \ModelicaSpeciesAmount{s_2}$, plus variables $\Stoich{s_1,R}$, $\Stoich{s_2,R}$, and $\Stoich{s_3,R}$ set to constant values $-1$, $-2$, and $+1$ respectively.

\subsubsection{The Event objects}
\label{sec:methods:encoding:event}
An \emph{Event} Modelica object is defined for each event $e$ defined in the input SBML model, in order to represent the event trigger condition, the event priority and delay (if any). All such objects are linked with the Model object.

In order to properly handle simultaneous events with conflicting assignments, the management of event assignments is split in three parts.

\emph{First}, an auxiliary variable is defined in the Event object encoding $e$ for each SBML model variable 
assigned by $e$.
When $e$ occurs (and after the event delay, if any), each such variable is set to the new value to be assigned to its associated model variable $v$, as stipulated by $e$.

\emph{Second}, conflicting assignments stemming from simultaneous events are resolved (within the Model object, Section~\ref{sec:methods:encoding:model}) by means of auxiliary functions, which also take into account the priorities of the competing events.

\emph{Third}, the objects owning the variables to be assigned after the occurring event(s) are informed of the final required changes and take care of actually performing the assignments.

Depending on whether each event is persistent or non-persistent, \Tool generates different code. Specialised and more efficient code is also generated for the common case of events with no delay.

\subsubsection{The Parameters object}
\label{sec:methods:encoding:parameters}
A single \emph{Parameters} Modelica object is defined to encode all the parameters of the input SBML model, together with their associated initial assignments and differential/algebraic equations stemming from SBML rate or assignments rules.

Hence, the Parameters object for Example~\ref{ex:methods:example} (which is linked with the Model object) would define and properly initialise variables $p_1$, $p_2$, $p_3$, $p_4$, and encode differential equations $\derive{p_2} = 0$ (stemming from rate rule $r_1$) and $\derive{p_4} = 0$ (stemming from the fact that value to $p_4$ changes only upon events).

\end{methods}

\section{Results}
\label{sec:results}
Section~\ref{sec:results:correctness} below shows the results of our experiments aimed at assessing the correctness of \Tool against the SBML Test Suite Core, while Section~\ref{sec:results:performance} compares the performance of general-purpose Modelica and FMU simulators when running the Modelica code generated by \Tool against the other SBML simulators for which an SBML Test Suite Core report is publicly available.

\subsection{Compliance to the SBML Test Suite Core}
\label{sec:results:correctness}
In order to assess the correctness of \Tool, we ran it against the test cases provided by SBML Test Suite Core v3.3.0, available in the SBML web-site. 

As \Tool aims at supporting the \emph{latest} SBML standard (\SBMLLatestStandard), we ignored the test cases involving \emph{deprecated} and \emph{discouraged} constructs such as \emph{fast reactions}. 
Hence, we ran \Tool against the remaining \TestSuiteNbIncluded out of the overall \TestSuiteNbAll test cases and simulated the generated Modelica code with the two major open-source Modelica implementations, namely OpenModelica (we used \OpenModelicaVersion) and JModelica (we used \JModelicaVersion), with the latter converting the input Modelica code into an FMU and then simulating such an FMU.

\begin{figure}
	\begin{center}
	\resizebox{\FigureStandardWidth}{!}{
		\begin{minipage}{0.4\linewidth}
		\renewcommand{\baselinestretch}{1.4}
		\setlength{\parindent}{0pt}
		\small
		\textbf{\Tool (\TestSuitePassPerc)}
		\par
		BioUML v2018.2 (100\%)
		\par
		COPASI v4.23.189 (74.26\%)
		\par
		Simul.\ Core Lib.\ v1.5 (70.04\%)
		\par
		iBioSIM v3.0.0 (70.04\%)
		\par
		libRoadRunner v1.4.18 (63.62\%)
		\par
		libSBMLSim v1.4.0 (61.30\%)
		\end{minipage}
		\begin{minipage}{0.5\linewidth}
			\includegraphics[width=\linewidth]{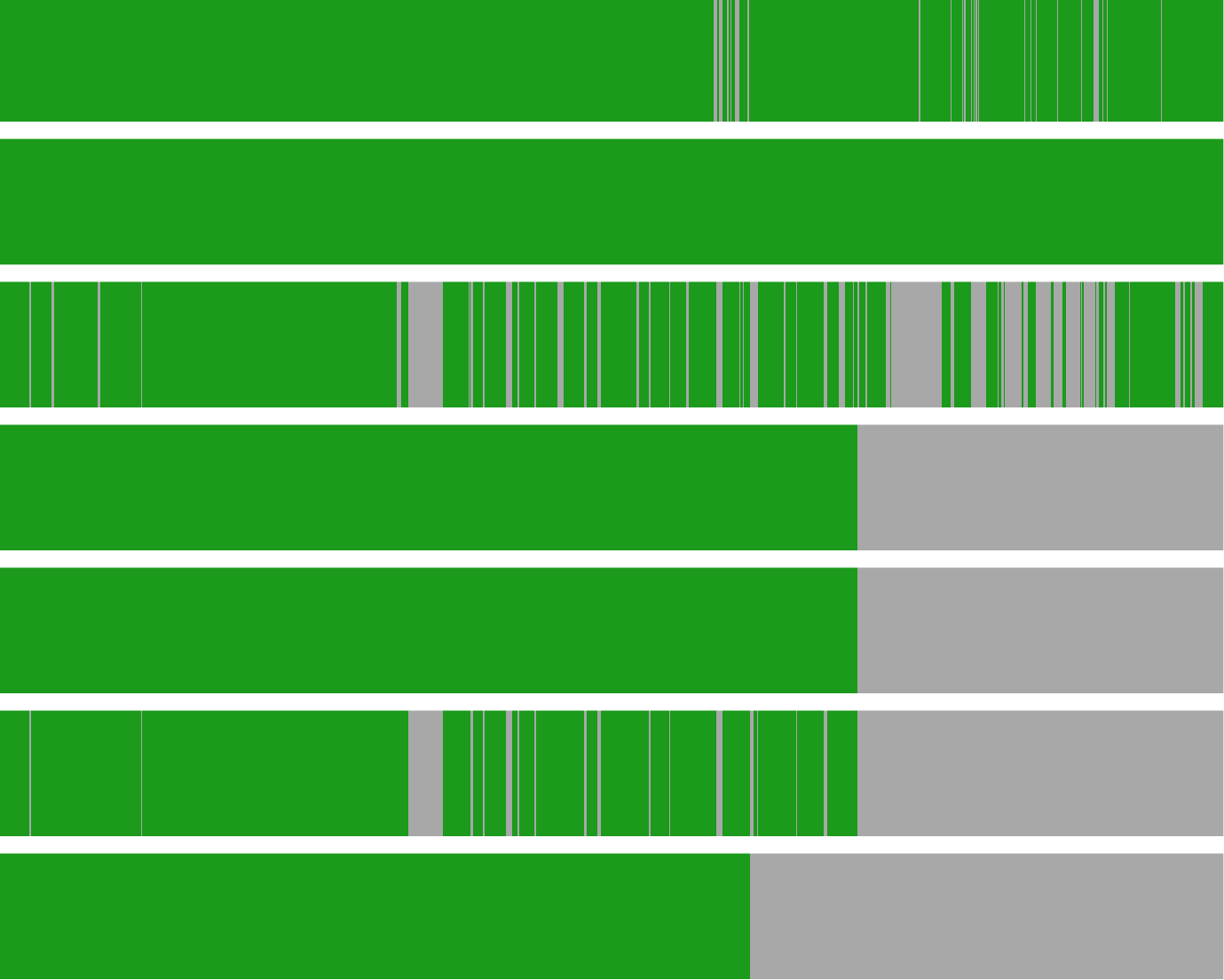}
		\end{minipage}
	}%
    \end{center}
    
	\caption{Compliance of \Tool-generated code (simulated by OpenModelica and JModelica) 
	to the SBML Test Suite Core v3.3.0, compared to other SBML simulators with certified public reports (test cases with deprecated SBML constructs are ignored).
    } 

    \label{fig:test-suite}
\end{figure}

\Tool achieves very high marks: the output (always identical between OpenModelica and JModelica) is exactly as expected on \TestSuiteNbPasses out of \TestSuiteNbIncluded test cases (\TestSuitePassPerc). 
Figure~\ref{fig:test-suite} compares the test cases of the SBML Test Suite Core v3.3.0 successfully simulated by OpenModelica/JModelica (on input provided by \Tool) to the results \emph{declared} by the other six systems for which a certified public report is available in the SBML web-site.
Namely, the figure shows, for each system, a series of \TestSuiteNbIncluded thin vertical bars, one per test case (sorted by their identifier in ascending order from left to right). 
Each vertical bar is coloured in green (respectively, grey) if the simulator output is (respectively, is not) exactly as expected by the SBML Test Suite Core upon numerical simulation of that test case.

The figure shows that \Tool{+}OpenModelica/JModelica rank among the top-compliant SBML simulators, being second only to BioUML.
%
Note that libSBMLSim, iBioSim, Simulation Core Library and libRoadRunner fail over a large number of test cases located on the right part of their plot. This is due to the fact that the test cases containing the constructs introduced in the latest SBML standard (not supported by them) have the highest values of their identifiers.

As for the \TestSuiteNbFails test cases for which the output computed by \Tool{+}OpenModelica/JModelica differs from the output expected by the test suite, 
in \TestSuiteNbFailsNotSignificant cases the difference is \emph{semantically meaningless}. In particular, the time series for the model variables computed by our Modelica simulators contain one more line with respect to the SBML Test Suite Core expected output, returning the value of the model variables immediately before each event (even if such events do not occur at time-points multiple of the requested sampling time). 
This is an intended behaviour of OpenModelica and JModelica, aimed at better showing the discontinuities in the values of the model variables that arise when events occur. 
By ignoring such additional lines, our output is exactly as expected.

The other \TestSuiteNbFailsUnsupportedCombs test cases where the output of \Tool{+}OpenModelica/JModelica differs from the output expected from the test suite are due to combinations of SBML constructs \emph{unsupported} by \Tool.  
Such combinations are discussed in Section~\ref{sec:discussion}. 
However, we anticipate that they are very rare in practice, semantically intricate, and easily avoidable.
Although some more involved/cryptic Modelica code could be in principle be generated to handle them, we chose to keep our output Modelica models as structured and human-intelligible as possible, in order to ease their extension and integration within larger model networks.

Anyway, there is no risk to accidentally generate flawed Modelica code, since any problematic combinations of constructs are \emph{statically detected} by \Tool, which warns the user during model translation about possible issues. 
The user can then act directly on the generated Modelica code to fix any raised issue.
Also, to further assist the user, suitable \emph{assertions} are injected in the generated Modelica code that would raise proper exceptions during simulation in case the Modelica model (generated \emph{with} warnings) actually behaves in a way not fully compliant to the SBML semantics.

\subsection{Model simulation performance}
\label{sec:results:performance}
In this section we aim at assessing to what extent translating SBML models into an open-standard general-purpose modelling language such as Modelica and into an open-standard general-purpose simulation ecosystem such as FMI/FMU introduces an overhead in simulation performance, when compared to simulation algorithms specialised to biochemical models.
To this end, we consider the major open-source (\emph{general-purpose}) Modelica and FMU simulators (OpenModelica and JModelica, respectively). Note that, while OpenModelica simulates the input model directly, JModelica works by translating the input Modelica model into an FMU and then by actually simulating such an FMU using the FMI API.

The performance of OpenModelica and JModelica/FMI in simulating SBML models translated by \Tool is compared against that of the \emph{specialised} SBML simulators reported in Section~\ref{sec:results:correctness} plus SystemModeler+BioChem, the only other available Modelica-based simulator for SBML models.

Although our results are by \emph{no means} to be intended as the outcome of a competition among SBML simulators, they clearly show that the advantages of using \Tool{+}OpenModelica and \Tool{+}JModelica/FMI \emph{do not generally come at any significant performance overhead}.

\subsubsection{Benchmarks}
\label{sec:results:performance:benchmarks}
We used the BioModels Database \cite{lenovere-etal:2006:biomodels}, a well-known repository of mathematical models of biological and biomedical systems taken from the scientific literature.
Models manually reviewed to guarantee reproducibility of results belong to the set of \emph{manually curated} models. This set of models is widely used as a benchmark for SBML interpreters and simulators.

We selected the subset of manually curated models of the BioModels Database (as of December 2018) that are \emph{accepted} by \Tool (\ie do not contain deprecated constructs or the unsupported combinations of constructs described in Section~\ref{sec:discussion}).
As a result, our benchmark set consists of \BioModelsNbIncluded models (out of the \BioModelsNbAll manually curated models of the BioModels Database), which have from \BioModelsNbVarsMin to \BioModelsNbVarsMax variables.

\subsubsection{Experimental setting}
\label{sec:results:performance:setting}
The computational complexity of a model simulation is affected by many factors, which are way beyond the mere number of variables. For example: number and structure of the differential as well algebraic equations; number and frequency of occurrence of events and complexity of their trigger conditions; algorithm and parameters used by the simulation engine. 

Setting up a methodologically-sound competition among OpenModelica, JModelica/FMI, and specialised SBML simulators is a complex task, which 
is clearly out of the scope of this paper.
So is the choice of the optimal simulation algorithm and configuration for a given model (in particular both OpenModelica and JModelica/FMI offer a wide portfolio of highly-configurable integrators to choose from).

Given our goals (see Section~\ref{sec:results:performance}), in our analysis we proceeded as follows.

\emph{First}, for each system we measured the \emph{core simulation time}, i.e., the time of simulating the given model from its (system-specific) \emph{internal representation}. 
This is consistent with the most demanding use-cases, such as parameter identification or estimation procedures, where simulator initialisation \emph{and} model preparation are performed only once, while a large number of simulations (with different parameter assignments) takes place, among which such initialisation costs are amortised.
As for OpenModelica and JModelica/FMI, this means that we ignored \Tool translation time, which, anyway, always takes \emph{less than 6\% of the simulation time}. 
For the same reason, as for JModelica/FMI we also ignored the time to generate the FMU (which is also \emph{negligible}). 

\emph{Second}, we used all systems with their \emph{default} integrators and settings. 

\emph{Third}, we fixed the simulation horizon and the maximum time step to, respectively, \BioModelsHorizon and \BioModelsMaxTimeStep (model) time units.
This last choice allowed us to extrapolate a clear performance trend of each system on the basis of the number of model variables.

All simulations were performed on a commodity computer (AMD A12-9720P CPU, 12 GB RAM, SSD, standard Linux environment), with a time-out of \BioModelsTimeout.

\subsubsection{Experimental results}
\label{sec:results:performance:results}

\begin{figure*}
    \begin{center}
        \includegraphics[width=\linewidth]{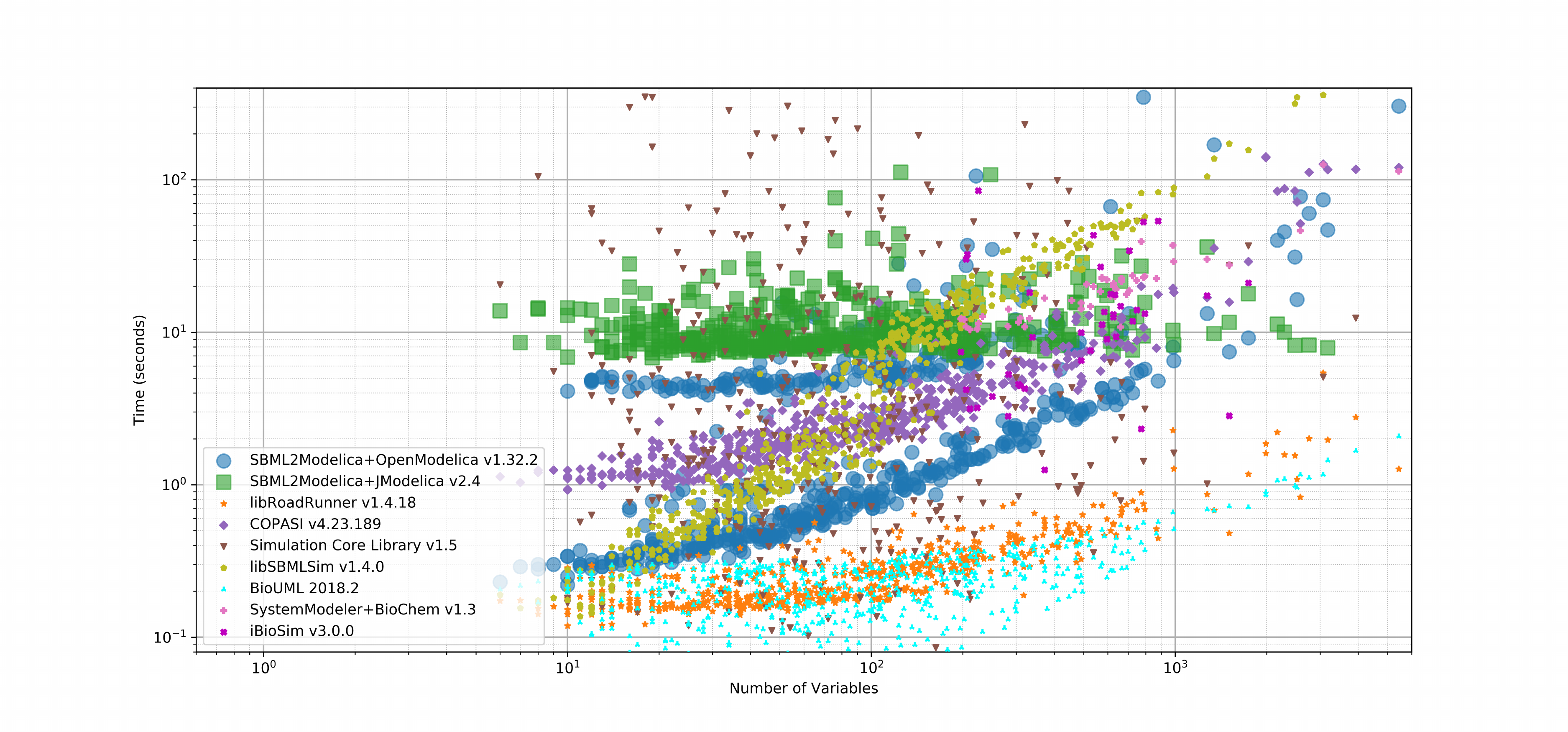}
        
        \caption{%
            Performance trend (log-log plot) of general-purpose OpenModelica (\OpenModelicaVersion) and JModelica/FMI (\JModelicaVersion) simulators on our benchmark models (from BioModels) translated with \Tool, compared with that of specialised SBML simulators.
            (Due to the need of manual GUI interaction, results of Wolfram SystemModeler+BioChem and iBioSim are reported only for models with at least 250 variables.)
        }
        \label{fig:performance}    
    \end{center}
\end{figure*}

The scatter (log-log) plot in Figure~\ref{fig:performance} shows a dot for each model in our benchmark set and each system (if that system terminated within our time-out). 
For each dot in position $(x,y)$, $x$ denotes the number of variables of the associated SBML model, while $y$ denotes the time (in seconds) required by the system associated to the dot colour to terminate.
SBML model import in Wolfram SystemModeler+BioChem and iBioSim requires manual user interaction via GUI and could not be automated. We overcame this issue by manually launching such systems on all models with \emph{at least} \BioModelsManualNbVarsMin variables.

Figure~\ref{fig:performance} shows that OpenModelica and JMod\-el\-ica/FMI are \emph{competitive} on most of the benchmark set, although, on the two largest models, there is a visible performance gap with respect to some of the other systems, with OpenModelica and JModelica going in time-out for, respectively, one and both of them. 
Also, OpenModelica appears generally faster than JModelica/FMI, even if it seems to slow down a bit when simulating models with frequently-occurring events (as shown by the bimodal behaviour on models having a similar number of variables). However, also in these cases, its overall performance remain aligned to that of several specialised SBML simulators. Conversely, performance of JModelica/FMI are more stable, regardless of the events occurrence frequency.

Besides the above peculiarities, the \emph{overall trend} emerging from Figure~\ref{fig:performance} is that the benefits enabled by a translation into a \emph{general-purpose} open-standard modelling language for dynamical systems such as Modelica and, as for JModelica/FMI, into a general-purpose open-standard simulation ecosystem such as FMI/FMU, generally come at \emph{no significant overhead}, when compared against performance of \emph{specialised} SBML simulators.
%


\section{Discussion}
\label{sec:discussion}
Section~\ref{sec:results} shows that, by translating SBML models into Modelica (and in turn, as for JModelica/FMI, into FMUs), \Tool effectively enables \emph{seamless integration} of SBML models into larger heterogeneous model networks without (in most cases) sacrificing simulation performance. 

Below we analyse the \TestSuiteNbFailsUnsupportedCombs test cases in the SBML Test Suite Core containing those combinations of SBML constructs not supported by \Tool (see Section~\ref{sec:results:correctness}).
Note that such construct combinations are \emph{rare} in real-world models, \emph{semantically intricate}, and can be \emph{easily circumvented}. 
\Tool always \emph{detects} such cases, and \emph{warns} the user accordingly, hence there is \emph{no risk} to run flawed Modelica code.

\subsection{Events recomputation}
\label{sec:discussion:unsupported:recomputation}
If, at a certain time instant $t$, \emph{multiple} events are \emph{simultaneously} triggered, such events are ordered by their priorities and the event $e$ with the \emph{highest} priority is executed first (Example~\ref{ex:methods:example} shows one such case). 
However, the execution of $e$ might change the value of variables that occur in the priority expression or in the trigger condition of some of the other events waiting to be executed.
In such situations (which can easily be circumvented by modelling the underlying mechanisms of such interfering events more explicitly), \Tool cannot generate correct Modelica code.
To avoid to generate a flawed translation, \Tool statically detects whether such interferences might occur between events and warns the user.

\subsection{Nested triggers}
\label{sec:discussion:unsupported:nested}
Assume that a persistent event $e$ is triggered for the first time at time $t_1$ and that it is requested to be delayed by duration $d_1$. 
If, for some reason, event $e$ is triggered again at time $t_2$ such that $t_1 < t_2 < t_1 + d_1$ and is requested to be delayed by duration $d_2$, the trigger must be recorded and $e$ must execute at time $t_2 + d_2$. 
As we cannot statically detect how many times the trigger of a persistent event can be nested, \Tool raises a warning during model translation, informing the user that a second trigger for a persistent model event might potentially occur when a previous trigger for the same event is on hold because of a delay. 
Also, \Tool injects an assertion into the generated Modelica code which stops the simulation in case this situation \emph{actually} occurs at run-time (making the model behaviour not compliant with the SBML semantics).
Again, such situations can be easily circumvented by modelling the underlying causes of event delays in more explicit ways.

\subsection{Negative time}
\label{sec:discussion:unsupported:negative-time}
SBML allows the definition of quantities also in \emph{negative} time-points. This may cause issues.
For example, the value of a model variable $x$ at time 0 could be re-assigned by an event trigger (processed at time 0) using an expression such as $\left(\pre{x}\right)(0)$. In this case, the SBML semantics stipulates that $\left(\pre{x}\right)(0)$ must be determined \emph{somewhere else} in the model (otherwise, a semantic error occurs).
Another example of negative time is the occurrence of $f(t - d)$, where $f$ is a function and $d > 0$ is the duration of a \emph{delay}. 
According to the SBML specifications, the value of $f(t - d)$ is defined also when $t < d$, by assuming that $f$ is defined also for negative time-points.
Conversely, the Modelica language specification forbids negative time-points, as time-point 0 is assumed to be the \emph{initial time-point} for simulation.

Although, in principle, additional Modelica code could be generated by \Tool in order to properly handle such cases (\eg by computing a suitable time-offset for all the model variables, and by artificially shifting in time the evolution of the entire model), we decided not to support this possibility, in order to keep the generated Modelica code well-structured and fully readable. 
Hence, when the above situations are detected, \Tool issues a warning. The user interested in supporting such cases, can take full responsibility by acting directly on the generated Modelica code.

\subsection{Unsupported math}
\label{sec:discussion:unsupported:math}
SBML allows users to explicitly assign value \texttt{NaN} to variables. When such values are found in the input SBML model, \Tool notices the user, since the numerical model simulation is clearly not possible.

\section{Conclusions}
\label{sec:conclusions}
In this paper we presented \Tool, an \SBMLLatestStandard--compliant software system that translates SBML models into well-structured, user-intelligible, easily modifiable \emph{Modelica} code, an open-standard general-purpose modelling language for which several efficient simulators (both commercial and open-source) are available. 
Modelica models can also be exported into black-box language-independent FMUs, an open standard supported by more than 100 simulators from virtually all application domains.

All this paves the way to the seamless integration (without lack of simulation performance) of SBML models within open-standard ecosystems, where biochemical models can be used as components of large heterogeneous model networks (together with models of, \eg human physiology, clinical protocol guidelines, treatment schemes, biomedical devices), and where standard system engineering approaches can be employed to perform their simulation-based analysis at system level.

\paragraph*{Acknowledgements.}
This work was partially supported by: 
Italian Ministry of University and Research under grant ``Dipartimenti di Eccellenza 2018--2022'' of the Department of Computer Science of Sapienza University of Rome; 
EC FP7 project PAEON (Model Driven Computation of Treatments for Infertility Related Endocrinological Diseases, 600773);
INdAM ``GNCS Project 2019'';
Sapienza University 2018 project RG11816436BD4F21 ``Computing Complete Cohorts of Virtual Phenotypes for In Silico Clinical Trials and Model-Based Precision Medicine''.
Authors are very grateful to the anonymous reviewers for their valuable comments.


\newcommand{\MCLabDOI}[1]{\nolinkurl{#1}}\newcommand{\MCLabProceedingsOf}[1]{Proceedings
  of #1}
\begin{thebibliography}{}

\bibitem[de~Bono and Hunter(2012)de~Bono and Hunter]{debono-etal:2012:biotech}
de~Bono, B. and Hunter, P. (2012).
\newblock Integrating knowledge representation and quantitative modelling in
  physiology.
\newblock {\em Biotechnology Journal\/}, {\bf 7}(8), 958--972.

\bibitem[Eissing {\em et~al.}(2011)Eissing, Kuepfer, Becker, Block, Coboeken,
  Gaub, Goerlitz, Jaeger, Loosen, Ludewig, Meyer, Niederalt, Sevestre,
  Siegmund, Solodenko, Thelen, Telle, Weiss, Wendl, Willmann, and
  Lippert]{eissing-etal:2011:frontiers}
Eissing, T., Kuepfer, L., Becker, C., Block, M., Coboeken, K., Gaub, T.,
  Goerlitz, L., Jaeger, J., Loosen, R., Ludewig, B., Meyer, M., Niederalt, C.,
  Sevestre, M., Siegmund, H., Solodenko, J., Thelen, K., Telle, U., Weiss, W.,
  Wendl, T., Willmann, S., and Lippert, J. (2011).
\newblock A computational systems biology software platform for multiscale
  modeling and simulation: Integrating whole-body physiology, disease biology,
  and molecular reaction networks.
\newblock {\em Frontiers in physiology\/}, {\bf 2}, 4.

\bibitem[Fritzson {\em et~al.}(2007)Fritzson, Ulfhielm, Belic, Fransson, and
  Gr{\`{e}}en]{fritzson-etal:2007:biochem}
Fritzson, P., Ulfhielm, E., Belic, A., Fransson, M., and Gr{\`{e}}en, H.
  (2007).
\newblock Biochemical mathematical modeling with modelica and the biochem
  library.
\newblock In {\em \MCLabProceedingsOf{6th International Conference
  APLIMAT~2007}\/}, pages 147--159.

\bibitem[Gendrault {\em et~al.}(2014)Gendrault, Madec, Lallement, and
  Haiech]{gendrault-etal:2014:ieee-tbe}
Gendrault, Y., Madec, M., Lallement, C., and Haiech, J. (2014).
\newblock Modeling biology with {HDL} languages: A first step toward a genetic
  design automation tool inspired from microelectronics.
\newblock {\em {IEEE} Transactions on Biomedical Engineering\/}, {\bf 61}(4),
  1231--1240.

\bibitem[Hucka {\em et~al.}(2003)Hucka, Finney, Sauro, Bolouri, Doyle, Kitano,
  Arkin, Bornstein, Bray, Cornish-Bowden, Cuellar, Dronov, Gilles, Ginkel, Gor,
  Goryanin, Hedley, Hodgman, Hofmeyr, Hunter, Juty, Kasberger, Kremling,
  Kummer, Le~Nov{\`{e}}re, Loew, Lucio, Mendes, Minch, Mjolsness, Nakayama,
  Nelson, Nielsen, Sakurada, Schaff, Shapiro, Shimizu, Spence, Stelling,
  Takahashi, Tomita, Wagner, and Wang]{hucka-etal:2003:bioinf}
Hucka, M., Finney, A., Sauro, H.~M., Bolouri, H., Doyle, J.~C., Kitano, H.,
  Arkin, A.~P., Bornstein, B.~J., Bray, D., Cornish-Bowden, A., Cuellar, A.~A.,
  Dronov, S., Gilles, E.~D., Ginkel, M., Gor, V., Goryanin, I.~I., Hedley,
  W.~J., Hodgman, T.~C., Hofmeyr, J.~H., Hunter, P.~J., Juty, N.~S., Kasberger,
  J.~L., Kremling, A., Kummer, U., Le~Nov{\`{e}}re, N., Loew, L.~M., Lucio, D.,
  Mendes, P., Minch, E., Mjolsness, E.~D., Nakayama, Y., Nelson, M.~R.,
  Nielsen, P.~F., Sakurada, T., Schaff, J.~C., Shapiro, B.~E., Shimizu, T.~S.,
  Spence, H.~D., Stelling, J., Takahashi, K., Tomita, M., Wagner, J., and Wang,
  J. (2003).
\newblock The {S}ystems {B}iology {M}arkup language ({SBML}): a medium for
  representation and exchange of biochemical network models.
\newblock {\em Bioinformatics\/}, {\bf 19}(4), 524--531.

\bibitem[Hucka {\em et~al.}(2018)Hucka, Bergmann, Dr{\"{a}}ger, Hoops, Keating,
  Le~Nov{\`{e}}re, Myers, Olivier, Sahle, Schaff, Smith, Waltemath, and
  Wilkinson]{hucka-etal:2018:sbml-l3v2}
Hucka, M., Bergmann, F., Dr{\"{a}}ger, A., Hoops, S., Keating, S.,
  Le~Nov{\`{e}}re, N., Myers, C., Olivier, B., Sahle, S., Schaff, J., Smith,
  L., Waltemath, D., and Wilkinson, D. (2018).
\newblock The {S}ystems {B}iology {M}arkup {L}anguage ({SBML}): {L}anguage
  specification for {L}evel 3 {V}ersion 2 {C}ore.
\newblock {\em Journal of Integrative Bioinformatics\/}, {\bf 15}(1).

\bibitem[Kanade {\em et~al.}(2009)Kanade, Alur, Ivancic, Ramesh,
  Sankaranarayanan, and Shashidhar]{kanade-etal:2009:cav}
Kanade, A., Alur, R., Ivancic, F., Ramesh, S., Sankaranarayanan, S., and
  Shashidhar, K. (2009).
\newblock Generating and analyzing symbolic traces of {S}imulink/{S}tateflow
  models.
\newblock In {\em \MCLabProceedingsOf{21st International Conference on Computer
  Aided Verification (CAV~2009)}\/}, volume 5643 of {\em Lecture Notes in
  Computer Science\/}, pages 430--445. Springer.

\bibitem[Keller {\em et~al.}(2013)Keller, D{\"{o}}rr, Tabira, Funahashi,
  Ziller, Adams, Rodriguez, Nov{\`{e}}re, Hiroi, Planatscher, Zell, and
  Dr{\"{a}}ger]{keller-etal:2013:simcorelib}
Keller, R., D{\"{o}}rr, A., Tabira, A., Funahashi, A., Ziller, M., Adams, R.,
  Rodriguez, N., Nov{\`{e}}re, N., Hiroi, N., Planatscher, H., Zell, A., and
  Dr{\"{a}}ger, A. (2013).
\newblock The systems biology simulation core algorithm.
\newblock {\em BMC Systems Biology\/}, {\bf 7}, 55.

\bibitem[Kolpakov {\em et~al.}(2019)Kolpakov, Akberdin, Kashapov, Kiselev,
  Kolmykov, Kondrakhin, Kutumova, Mandrik, Pintus, Ryabova, Sharipov, Yevshin,
  and Kel]{kolpakov:2019:biouml}
Kolpakov, F., Akberdin, I., Kashapov, T., Kiselev, L., Kolmykov, S.,
  Kondrakhin, Y., Kutumova, E., Mandrik, N., Pintus, S., Ryabova, A., Sharipov,
  R., Yevshin, I., and Kel, A. (2019).
\newblock {BioUML}: an integrated environment for systems biology and
  collaborative analysis of biomedical data.
\newblock {\em Nucleic Acids Research\/}, {\bf 47}(W1), W225--W233.

\bibitem[Larsdotter~Nilsson and Fritzson(2003)Larsdotter~Nilsson and
  Fritzson]{larsdotter-etal:2003:biochem}
Larsdotter~Nilsson, E. and Fritzson, P. (2003).
\newblock {BioChem} - {A} biological and chemical library for {M}odelica.
\newblock In {\em \MCLabProceedingsOf{3rd International Modelica Conference
  (Modelica~2003)}\/}, pages 215--220.

\bibitem[Le~Nov{\`{e}}re {\em et~al.}(2006)Le~Nov{\`{e}}re, Bornstein,
  Broicher, Courtot, Donizelli, Dharuri, Li, Sauro, Schilstra, Shapiro, Snoep,
  and Hucka]{lenovere-etal:2006:biomodels}
Le~Nov{\`{e}}re, N., Bornstein, B., Broicher, A., Courtot, M., Donizelli, M.,
  Dharuri, H., Li, L., Sauro, H., Schilstra, M., Shapiro, B., Snoep, J., and
  Hucka, M. (2006).
\newblock {BioModels Database}: {A} free, centralized database of curated,
  published, quantitative kinetic models of biochemical and cellular systems.
\newblock {\em Nucleic Acids Research\/}, {\bf 34}(suppl\_1), D689--D691.

\bibitem[Lee {\em et~al.}(2006)Lee, Xu, Singhal, Mendes, Hoops, \`{u}~Pahle,
  Simus, Gauges, Sahle, and Kummer]{lee-etal:2006:copasi}
Lee, C., Xu, L., Singhal, M., Mendes, P., Hoops, S., \`{u}~Pahle, J., Simus,
  N., Gauges, R., Sahle, S., and Kummer, U. (2006).
\newblock {COPASI} ---- {A} {CO}mplex {PA}thway {SI}mulator.
\newblock {\em Bioinformatics\/}, {\bf 22}(24), 3067--3074.

\bibitem[Madec {\em et~al.}(2017)Madec, Lallement, and
  Haiech]{madec-etal:2017:bbspice}
Madec, M., Lallement, C., and Haiech, J. (2017).
\newblock Modeling and simulation of biological systems using {SPICE} language.
\newblock {\em PLoS ONE\/}, {\bf 12}(8), 1--21.

\bibitem[Madec {\em et~al.}(2019)Madec, Bonament, Rosati, Takakura, Haiech,
  H{\'{e}}brard, and Lallement]{madec-etal:2019:lab-on-chips}
Madec, M., Bonament, A., Rosati, E., Takakura, Y., Haiech, J., H{\'{e}}brard,
  L., and Lallement, C. (2019).
\newblock Environment for modeling and simulation of biosystems, biosensors,
  and lab-on-chips.
\newblock {\em {IEEE} Transactions on Electron Devices\/}, {\bf 66}(1), 34--43.

\bibitem[Mancini {\em et~al.}(2013)Mancini, Mari, Massini, Melatti, Merli, and
  Tronci]{mancini-etal:2013:cav}
Mancini, T., Mari, F., Massini, A., Melatti, I., Merli, F., and Tronci, E.
  (2013).
\newblock System level formal verification via model checking driven
  simulation.
\newblock In {\em \MCLabProceedingsOf{25th International Conference on Computer
  Aided Verification (CAV~2013)}\/}, volume 8044 of {\em Lecture Notes in
  Computer Science\/}, pages 296--312. Springer.

\bibitem[Mancini {\em et~al.}(2014)Mancini, Mari, Massini, Melatti, and
  Tronci]{mancini-etal:2014:pdp}
Mancini, T., Mari, F., Massini, A., Melatti, I., and Tronci, E. (2014).
\newblock System level formal verification via distributed multi-core hardware
  in the loop simulation.
\newblock In {\em \MCLabProceedingsOf{22nd Euromicro International Conference
  on Parallel, Distributed, and Network-Based Processing (PDP~2014)}\/}, pages
  734--742. IEEE.

\bibitem[Mancini {\em et~al.}(2015)Mancini, Tronci, Salvo, Mari, Massini, and
  Melatti]{mancini-etal:2015:iwbbio}
Mancini, T., Tronci, E., Salvo, I., Mari, F., Massini, A., and Melatti, I.
  (2015).
\newblock Computing biological model parameters by parallel statistical model
  checking.
\newblock In {\em \MCLabProceedingsOf{3rd International Conference on
  Bioinformatics and Biomedical Engineering (IWBBIO~2015)}\/}, volume 9044 of
  {\em Lecture Notes in Computer Science\/}, pages 542--554. Springer.

\bibitem[Mancini {\em et~al.}(2016a)Mancini, Mari, Massini, Melatti, and
  Tronci]{mancini-etal:2016:micpro}
Mancini, T., Mari, F., Massini, A., Melatti, I., and Tronci, E. (2016a).
\newblock Anytime system level verification via parallel random exhaustive
  hardware in the loop simulation.
\newblock {\em Microprocessors and Microsystems\/}, {\bf 41}, 12--28.

\bibitem[Mancini {\em et~al.}(2016b)Mancini, Mari, Massini, Melatti, and
  Tronci]{mancini-etal:2016:fundam}
Mancini, T., Mari, F., Massini, A., Melatti, I., and Tronci, E. (2016b).
\newblock {SyLVaaS}: System level formal verification as a service.
\newblock {\em Fundamenta Informaticae\/}, {\bf 1--2}, 101--132.

\bibitem[Mancini {\em et~al.}(2017)Mancini, Mari, Massini, Melatti, Salvo, and
  Tronci]{mancini-etal:2017:ipl}
Mancini, T., Mari, F., Massini, A., Melatti, I., Salvo, I., and Tronci, E.
  (2017).
\newblock On minimising the maximum expected verification time.
\newblock {\em Information Processing Letters\/}, {\bf 122}, 8--16.

\bibitem[Mancini {\em et~al.}(2018)Mancini, Mari, Massini, Melatti, Salvo,
  Sinisi, Tronci, Ehrig, R{\"{o}}blitz, and
  Leeners]{mancini-etal:2018:rcra-treatment}
Mancini, T., Mari, F., Massini, A., Melatti, I., Salvo, I., Sinisi, S., Tronci,
  E., Ehrig, R., R{\"{o}}blitz, S., and Leeners, B. (2018).
\newblock Computing personalised treatments through in silico clinical trials.
  {{A}} case study on downregulation in assisted reproduction.
\newblock In {\em \MCLabProceedingsOf{25th RCRA International Workshop on
  Experimental Evaluation of Algorithms for Solving Problems with Combinatorial
  Explosion (RCRA~2018)}\/}, volume 2271 of {\em CEUR Workshop Proceedings\/}.
  CEUR.org.

\bibitem[Matej{\'{a}}k and Kofr{\'{a}}nek(2015)Matej{\'{a}}k and
  Kofr{\'{a}}nek]{matejak-etal:2015:embc}
Matej{\'{a}}k, M. and Kofr{\'{a}}nek, J. (2015).
\newblock Physiomodel -- {A}n integrative physiology in {M}odelica.
\newblock In {\em \MCLabProceedingsOf{37th Annual International Conference of
  the IEEE Engineering in Medicine and Biology Society (EMBC~2015)}\/}, pages
  1464--1467. IEEE.

\bibitem[Messori {\em et~al.}(2018)Messori, Incremona, Cobelli, and
  Magni]{messori-etal:2018:individualized}
Messori, M., Incremona, G., Cobelli, C., and Magni, L. (2018).
\newblock Individualized model predictive control for the artificial pancreas:
  In silico evaluation of closed-loop glucose control.
\newblock {\em {IEEE} Control Systems Magazine\/}, {\bf 38}(1), 86--104.

\bibitem[Myers {\em et~al.}(2009)Myers, Barker, Jones, Kuwahara, Madsen, and
  Nguyen]{myers-etal:2009:ibiosim}
Myers, C., Barker, N., Jones, K., Kuwahara, H., Madsen, C., and Nguyen, N.-P.
  (2009).
\newblock {iBioSim}: {A} tool for the analysis and design of genetic circuits.
\newblock {\em Bioinformatics\/}, {\bf 25}(21), 2848--2849.

\bibitem[Schaller {\em et~al.}(2016)Schaller, Lippert, Schaupp, Pieber,
  Schuppert, and Eissing]{schaller-etal:2016:mpc}
Schaller, S., Lippert, J., Schaupp, L., Pieber, T., Schuppert, A., and Eissing,
  T. (2016).
\newblock Robust {PBPK/PD}-based model predictive control of blood glucose.
\newblock {\em {IEEE} Transactions on Biomedical Engineering\/}, {\bf 63}(7),
  1492--1504.

\bibitem[Somogyi {\em et~al.}(2015)Somogyi, Bouteiller, Glazier, K{\"{o}}nig,
  Medley, Swat, and Sauro]{somogyi-etal:2015:libroadrunner}
Somogyi, E., Bouteiller, J.-M., Glazier, J., K{\"{o}}nig, M., Medley, J., Swat,
  M., and Sauro, H. (2015).
\newblock {libRoadRunner}: {A} high performance {SBML} simulation and analysis
  library.
\newblock {\em Bioinformatics\/}, {\bf 31}(20), 3315--3321.

\bibitem[Takizawa {\em et~al.}(2013)Takizawa, Nakamura, Tabira, Chikahara,
  Matsui, Hiroi, and Funahashi]{takizawa-etal:2013:libsbmlsim}
Takizawa, H., Nakamura, K., Tabira, A., Chikahara, Y., Matsui, T., Hiroi, N.,
  and Funahashi, A. (2013).
\newblock {LibSBMLSim}: {A} reference implementation of fully functional {SBML}
  simulator.
\newblock {\em Bioinformatics\/}, {\bf 29}(11), 1474--1476.

\bibitem[Tronci {\em et~al.}(2014)Tronci, Mancini, Salvo, Sinisi, Mari,
  Melatti, Massini, Dav{i'}, Dierkes, Ehrig, R{\"{o}}blitz, Leeners,
  Kr{\"{u}}ger, Egli, and Ille]{mancini-etal:2014:fmcad}
Tronci, E., Mancini, T., Salvo, I., Sinisi, S., Mari, F., Melatti, I., Massini,
  A., Dav{i'}, F., Dierkes, T., Ehrig, R., R{\"{o}}blitz, S., Leeners, B.,
  Kr{\"{u}}ger, T.~H.~C., Egli, M., and Ille, F. (2014).
\newblock Patient-specific models from inter-patient biological models and
  clinical records.
\newblock In {\em \MCLabProceedingsOf{14th International Conference on Formal
  Methods in Computer-Aided Design (FMCAD 2014)}\/}, pages 207--214. IEEE.

\bibitem[Zuliani(2015)Zuliani]{zuliani:2015:jttt}
Zuliani, P. (2015).
\newblock Statistical model checking for biological applications.
\newblock {\em International Journal on Software Tools for Technology
  Transfer\/}, {\bf 17}(4), 527--536.

\bibitem[Zuliani {\em et~al.}(2013)Zuliani, Platzer, and
  Clarke]{zuliani-etal:2013:fmsd}
Zuliani, P., Platzer, A., and Clarke, E. (2013).
\newblock {B}ayesian statistical model checking with application to
  {S}tateflow/{S}imulink verification.
\newblock {\em Formal Methods in System Design\/}, {\bf 43}(2), 338--367.

\end{thebibliography}

\end{document}